\documentclass{aa}
\usepackage{txfonts,graphicx}

\long\def\jumpover#1{{}}

\def \th{\thinspace}

\def \Teff{{$T_{\rm{ef\!f}} $}}
\def \Mo{{$M_\odot $}}

\def \apriori{{\it a priori\ }}

\def \eg{{{\it e.g.},\ }}
\def \etal{{et al.}}

\def \cf{{\it cf.\ }}

\def \ie{{{\it i.e.},\ }}

\def \viz{{\it viz.\ }}
\def \vs{{\it vs.\ }}

\def\approxgt{\,\raise2pt \hbox{$>$}\kern-8pt\lower2.pt\hbox{$\sim$}\,\th}
\def\approxlt{\,\raise2pt \hbox{$<$}\kern-8pt\lower2.pt\hbox{$\sim$}\,\th}
\def\v{{\rm v}}

\begin{document}

\title{Magellanic Cloud Cepheids: Pulsational and Evolutionary Modelling vs.
           Observations}
\titlerunning{MC Cepheids}
\authorrunning{Buchler et al.}
\author{J. Robert Buchler\inst{1}
        \and Zolt\'an Koll\'ath\inst{2}
        \and Jean-Philippe Beaulieu\inst{3}}
\institute{
Physics Department, University of Florida, Gainesville, FL 32611, USA,\quad
\email{buchler@phys.ufl.edu}
\and
Konkoly Observatory, Budapest, HUNGARY,\quad
\email{kollath@konkoly.hu}
\and
Institut d'Astrophysique de Paris, FRANCE,\quad
\email{beaulieu@iap.fr}
}
\offprints{buchler@phys.ufl.edu}
\date{submitted January 2004, \th  accepted May 7, 2004}

\abstract{The pulsational properties of the Cepheid models along the
evolutionary tracks from the Padova group (Girardi \etal), as calculated with
our turbulent convective pulsation code, are in good agreement with the
resonance constraints imposed by the observational OGLE-2 data of the Small and
Large Magellanic Clouds.  Our study suggests that the $P_4/P_1=1/2$ resonance
for the overtone Cepheids occurs for periods clustering around 4.2\th d, in
disagreement with the suggestion of Antonello \& Poretti based on the
observations of light curves, but in agreement with Kienzle \etal\ and
Feuchtinger \etal.  For the fundamental Cepheids the lowest order Fourier
decomposition coefficients from the light curves, \viz $R_{21}$ and $\phi_{21}$
can be used to locate the resonance region, but not so for the first overtone
Cepheids.  Here, the radial velocity curves can be used to locate the overtone
resonance region, or in their absence, one needs to resort to numerical
hydrodynamic modelling.   }

\maketitle


\section{Introduction}

Cepheids are perhaps the best observed variable stars.  They are very
interesting heat engines from a physical point of view, and it is hardly
necessary to stress their importance as cosmological distance indicators.
Obviously, a good theoretical understanding of these important stars is
essential, and they have received the most theoretical attention among the
variable stars, going as far back as a half-century.

It was first noted by Hertzsprung (1926) that the shape of the light curves
(LC) of the Galactic fundamental (F) mode Cepheids exhibit a progression with
pulsation period in the vicinity of $10$\th days.  Subsequently Payne-Gaposhkin
(1947) found concomitant sharp variations in the Fourier decomposition
coefficients of the LCs.  Later, Simon \& Schmidt (1976) noted a correlation of
the location of the bump on the LC with the location of a resonance between the
excited fundamental pulsation mode and the second overtone ($P_2/P_0 = 1/2$).

Intrigued by these features Buchler \& Goupil (1984) developed the mathematical
'amplitude equation' formalism that is necessary to understand how the
occurrence of an internal resonance can produce, both qualitatively and
quantitatively, a progression of the Fourier coefficients of the LCs and of the
radial velocity ($V_{r}$) curves (see also Buchler 1993 and Buchler \& Kov\'acs,
1986).  The amplitude equations for a given star yield the modal amplitudes as
a function of the stellar parameters, such as the period or \Teff\ for example,
and these in turn are related to the Fourier coefficients of the LC.

\begin{figure*}
 \centerline{\resizebox{8.9cm}{!}{\includegraphics{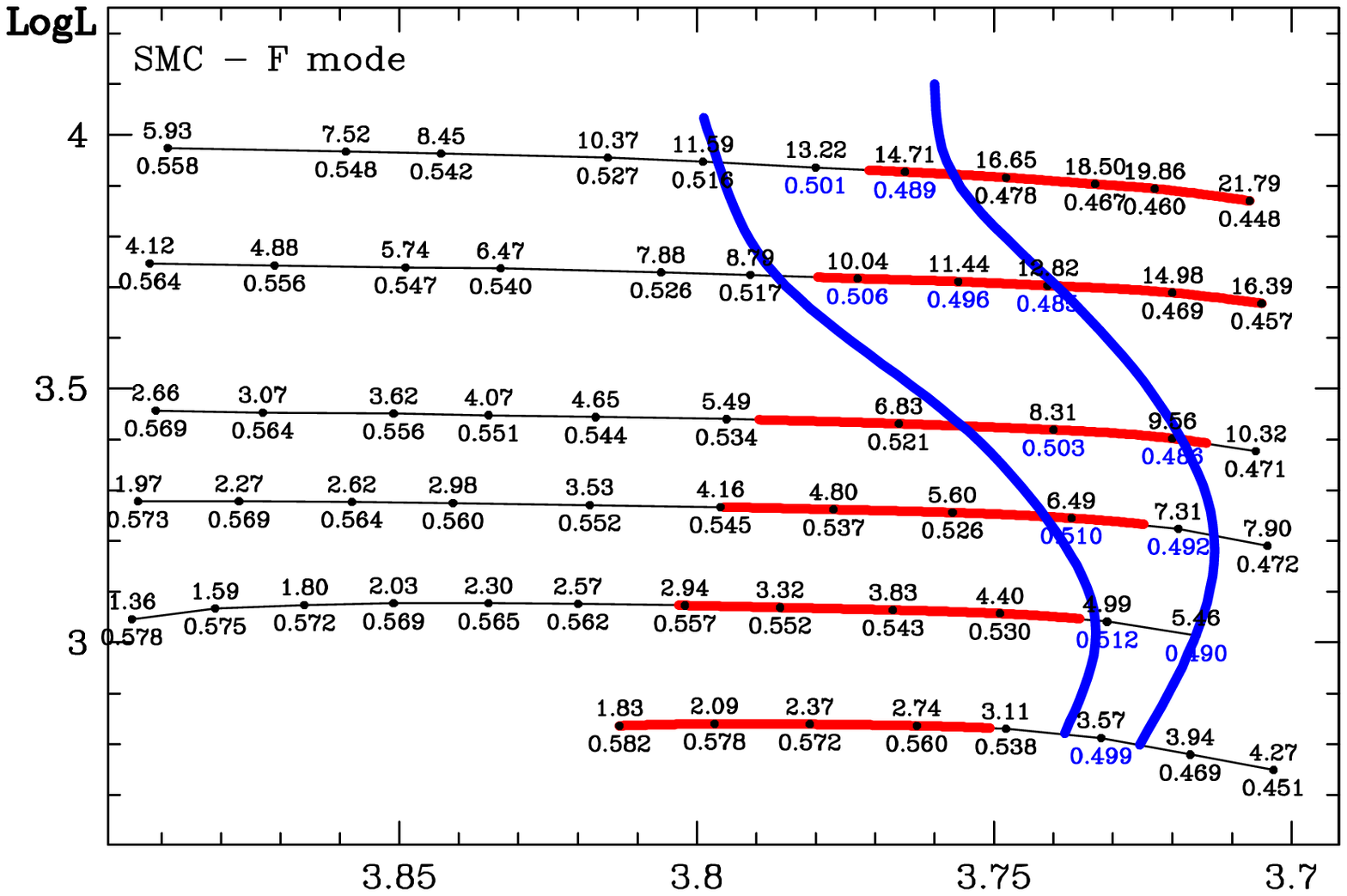}}
             \hskip 4pt
             \resizebox{8.9cm}{!}{\includegraphics{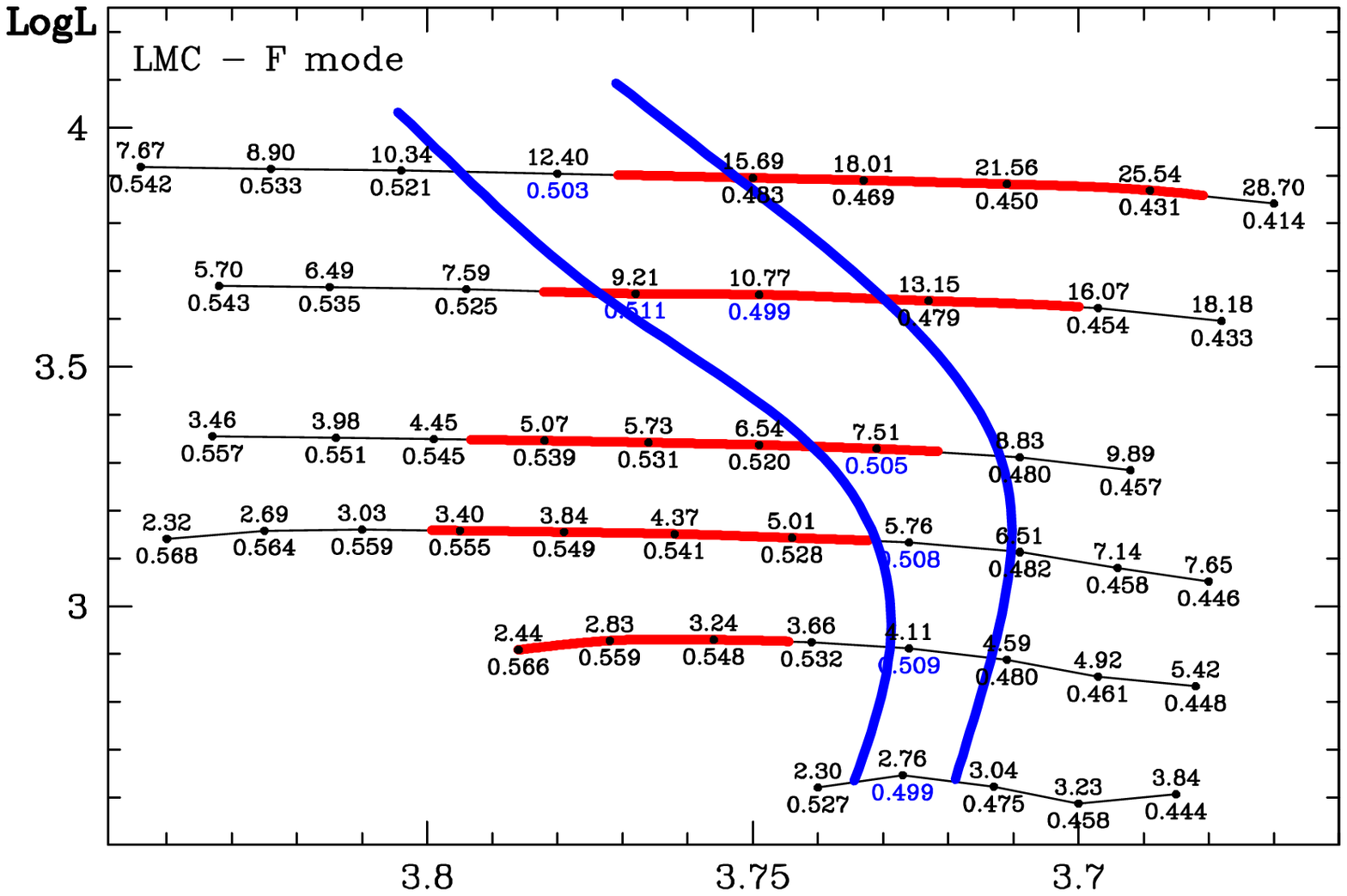}}}
  \vskip 3pt
 \centerline{\resizebox{8.9cm}{!}{\includegraphics{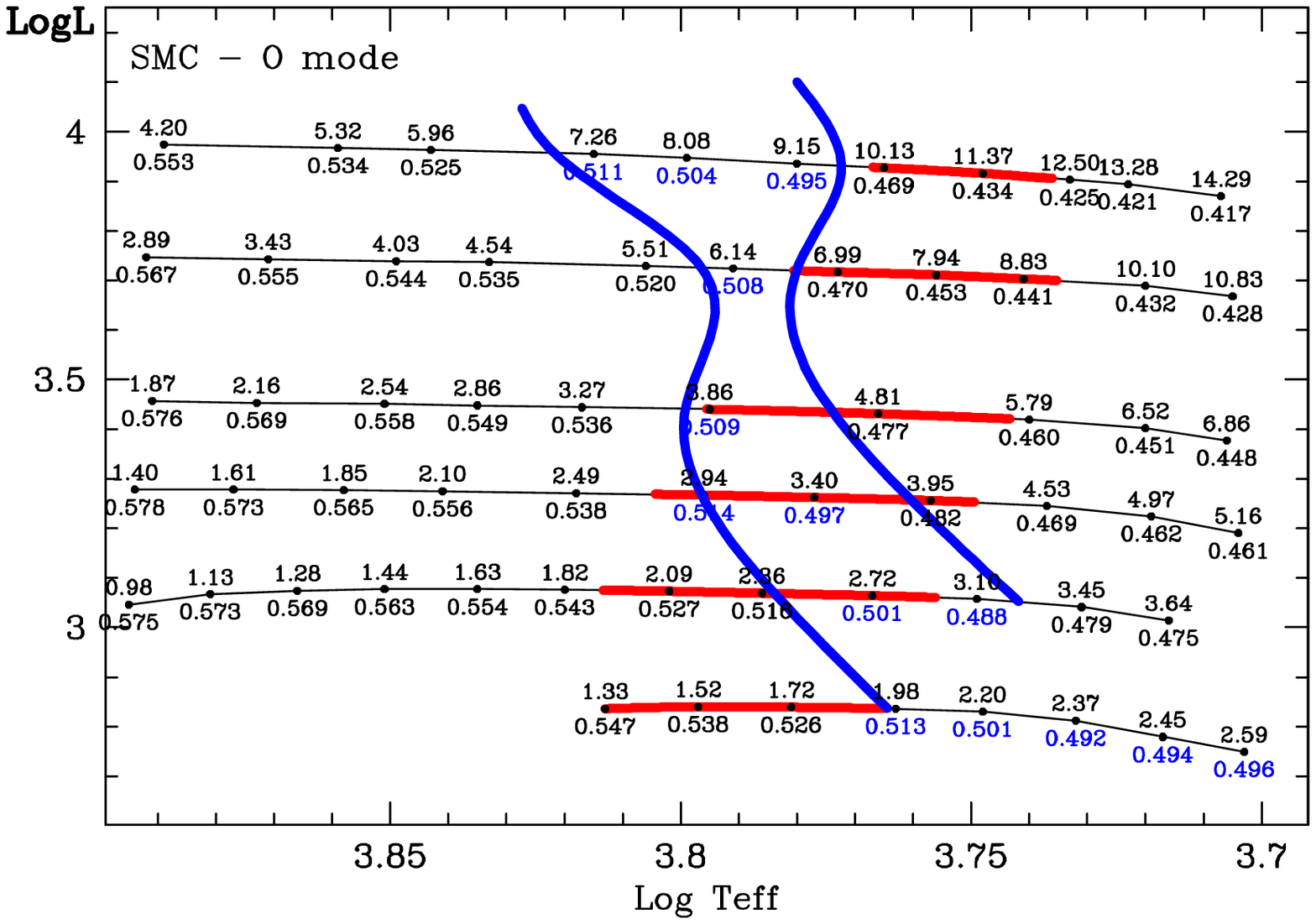}}
             \hskip 4pt
             \resizebox{8.9cm}{!}{\includegraphics{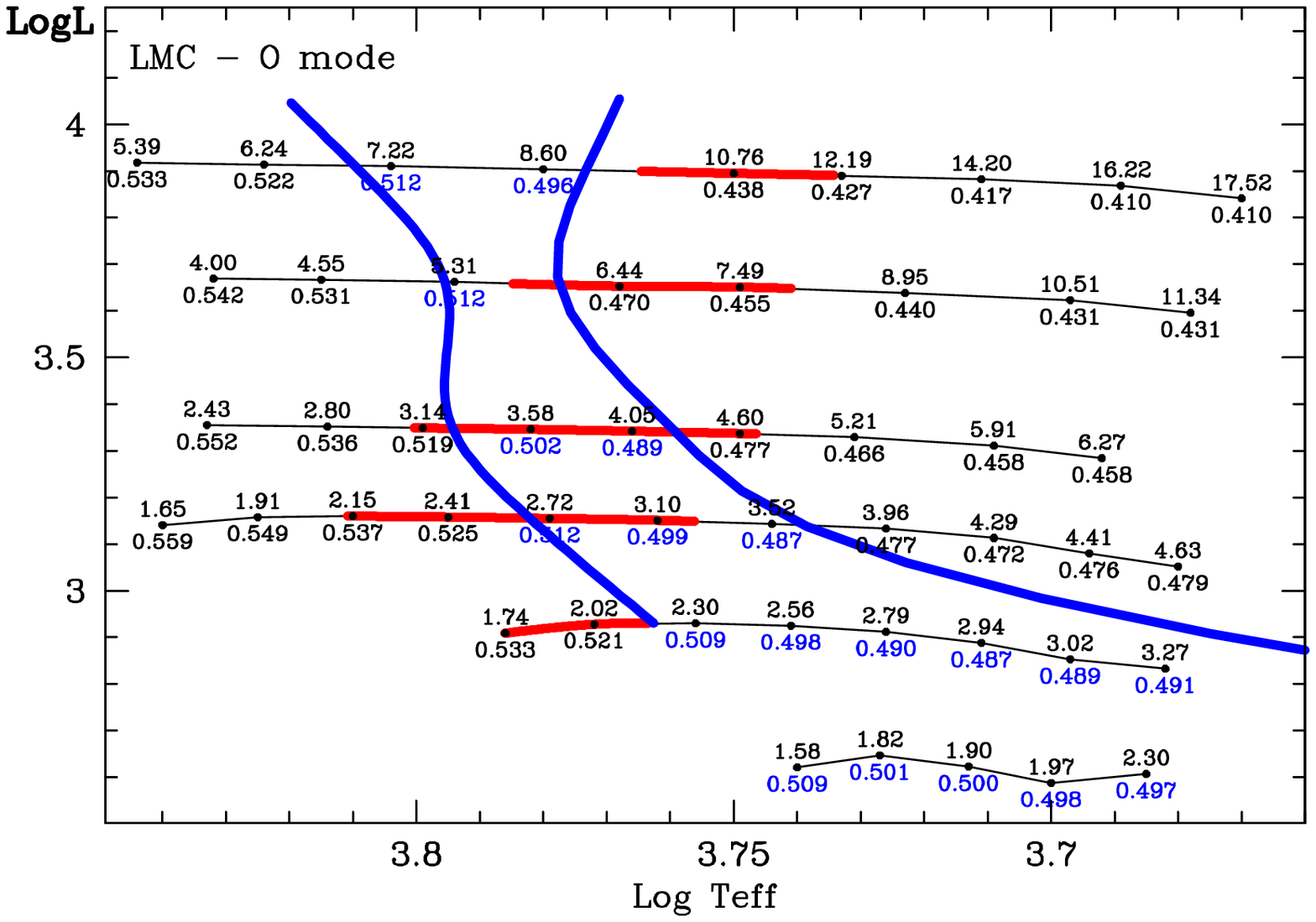}}}
 \parbox[b]{18cm}{
  \caption[]{\small {\bf F and O1 Cepheid models in SMC and LMC)}: 
    Sections of the 3.5, 4.0, 4.5, 5.0, 6.0 and 7.0 \Mo\ Padova 
    ($Z$=0.004 and 0.008) tracks corresponding to the 
    third crossing of the IS; 
    the periods [d] are noted above the tracks, 
    the period ratios $P_r$ below; the {\sl thickened sections} of the tracks
    represent the linear IS (linearly unstable
    models);  the pairs of thick {\sl vertical curves}
    delimit the resonance regions,
    $P_r = 0.5 \pm 0.015$.}
  \label{ber}
 }
 \end{figure*}

Klapp, Buchler \& Goupil (1985) and Kov\'acs \& Buchler (1989) applied this
formalism to the specific case of the F Cepheid progression.  A comparison of a
number of sequences of full amplitude hydrodynamic Cepheid models in Figs.~3, 5
and 7 of Buchler, Moskalik \& Kov\'acs (1990) shows very clearly that the
structure of the (magnitude) Fourier coefficients $\phi^m_{21}=
\phi^m_2-2\phi^m_1$ and $R^m_{12}= A_2/A_1$ correlates very well with the
period ratio $P_r=P_2/P_0$, which is indicative of the resonance, rather that
with the period $P_0$ itself.  (We note that it is because of the finite width
of the IS that stars with different masses and luminosities, and thus different
periods can have the same $P_r$ as Fig.~\ref{ber} below will show.  A corollary
is that there is no real Cepheid resonance center, but rather a resonance
region).  For the observed Cepheids we know the periods, but we do not have any
direct information on the period ratios $P_{r}$.  We therefore try to take
advantage of the behavior of the Fourier coefficients to localize the resonance
region.

The Fourier coefficients $\phi^{\v}_{21}$ and $R^{\v}_{21}$ of the observed
$V_{r}$ curves of Galactic F Cepheids also show structure in the vicinity of a
10\th d period (\eg Kov\'acs, Kisvarsanyi \& Buchler 1990), but the structure
is different from that of the LCs.  This is to be expected because the
luminosity $L\sim R^2 T^4$ involves not just the radial displacement
eigenvectors but also the temperature eigenvectors.  The coefficients in the
transformation from the modal amplitudes to LCs and to $V_{r}$ curves are
therefore different as explained in Kov\'acs \& Buchler (1989).

Thus, for example, in the F Cepheids the sharp feature in the magnitude phases
$\phi^{m}_{21}$ occurs close to 10\th d, and $R^{m}_{21}$ has a minimum there
\th~(\cf \eg Fig.~6 of Moskalik, Buchler \& Marom 1992).  The corresponding
$V_{r}$ phase $\phi^{\v}_{21}$ has a rapid, but not sharp variation from 5 to
15\th d, while the minimum for $R^{\v}_{21}$ is less pronounced and appears at
a somewhat larger period of $\sim$ 12\th d.  Either one of the two
coefficients, $\phi^{m}_{21}$ or $R^{m}_{21}$, can therefore serve as a
criterion for locating the resonance, but $\phi^{\v}_{21}$ and $R^{\v}_{21}$
are less useful discriminants.


We now turn to the O1 Cepheids.  Antonello \& Poretti (1986) suggested evidence
for an internal resonance of the excited first overtone with the fourth
overtone $P_{r}=P_4/P_1\sim 1/2$ in the vicinity of $P_1\sim 3.2$\th d, on the
basis of the sharp variations of $\phi^{m}_{21}$ of the Galactic LC Fourier
coefficients.  Linear models of O1 Cepheids indeed display such a resonance in
that general vicinity (Antonello 1994, Buchler \etal, 1996), the exact location
depending however on the $M$ -- $L$ relation that is used.  The first results
of nonlinear radiative model did not correctly reproduce the variation of the
Fourier coefficients (Antonello \& Aikawa 1995).  In contrast, on the basis of
the $V_{r}$ Fourier coefficients Kienzle, Moskalik, Bersier \& Pont (1999),
instead, put the resonance near 4.6\th d.  The reason for this discrepant
conclusion is that the Fourier coefficients, especially $\phi^m_{21}$ have a
substantially different behavior for the LCs and the $V_{r}$ curves.  See \eg
Feuchtinger, Buchler \& Koll\'ath (2000, hereafter FBK) for a juxtaposition.

In order to resolve this discrepancy it was necessary to make a detailed
numerical hydrodynamical survey of full amplitude pulsations (FBK).  The code
contains a time-dependent mixing length model for convection because purely
radiative models had failed to reproduce the observed features of the Fourier
coefficients, in particular near 3.2\th d.  This survey clearly put the
resonance  near 4.2\th d.  But it also showed that only the broad maximum
of $\phi^{\v}_{21}$ correlates with the resonance.

The conclusion is that while it is true that sharp features in the Fourier
coefficients generally indicate the presence of internal resonances, there is
{\sl no simple} \apriori criterion for locating precisely the resonance region
from either the LC or from the $V_{r}$ data.  Numerical hydrodynamic modelling
of full amplitude pulsations is needed to determine the criterion to be used
and the location of the resonance.

The Magellanic Cloud F Cepheid LCs appear very similar to those of the Galactic
ones in terms of their Fourier coefficients (\eg Beaulieu et al., 1995,
Beaulieu \& Sasselov 1997).  What matters in particular for this paper is that
the resonant features are very similar.  The observed LC features of the LMC
and SMC F Cepheids occur in the same place as for the Galactic ones.  A
systematic and comparative hydrodynamic full amplitude survey of Galactic and
MC F Cepheids is still lacking, and in its absence we make the reasonable
assumption that the Fourier structure as a function of $P_r$ is the same.

There are some small differences though for the overtones.  Despite a fair
amount of scatter, it would appear that the prominent sharp features, namely
the minimum in $R^{m}_{12}$ occurs near 2.7 and 2.4\th d for the LMC and SMC O1
Cepheids respectively (Beaulieu et al., 1995, Beaulieu \& Sasselov 1997, and
the larger OGLE sample by Udalski \etal 1999a, 1999b), whereas they occur
closer to 3.2\th d for the Galactic O1 Cepheids Antonello \& Poretti (1986).
There is a similar small shift in the sharp drop in $\phi^m_{21}$ to lower
periods.  In the absence of a numerical hydrodynamic survey similar to that of
FBK, we will therefore assume on the basis of these shifts in $R^m_{21}$ that
there are probably corresponding shifts of --0.5 and --0.7\th d of the
resonance period for the LMC and SMC O1 Cepheids.

In a previous attempt to use the resonances to obtain constraints on the $M$
and $L$ of the Cepheid models Buchler, Koll\'ath, Beaulieu \& Goupil (1996)
(see also Simon \& Kanbur 1994, Aikawa \& Antonello 2000) made the assumption
that the resonance boundaries, for example for F Cepheids, were defined for the
resonance center, \viz $P_2/P_0$= 1/2, at $P_0$ = 11\th d at the blue edge, and
9\th d at the red edge of the IS.  This led to their suggestion that the
derived masses for the SMC and LMC were too small.  The current paper
reexamines this assumption.  Fig.\ref{ber} and the discussion below show that
instead one should define the resonance boundaries to be
$P_r=P_2/P_0=0.5+\Delta$ at the blue edge and $P_r=0.5-\Delta$ at the red edge,
with $\Delta = 0.015$ for example, both for $P_0$ = 10\th d and $P_1=4.2$\th d
for O1 Cepheids).  If one repeats their procedure with these new conditions one
finds that the sensitivity to the exact values of the resonance boundaries
($P_r$) is too great to constrain $M$ and $L$.

Hence we are led here to an alternate comparison, still involving the
resonances, that is based on a combination of stellar evolution tracks and
pulsation properties.  Other studies (Baraffe et al., 1998, Alibert et al.,
1999, Bono et al., 2000a, Bono et al., 2000b, Bono et al., 2001) usually
concentrate on the comparison of the models with the observations in the period
-- luminosity planes in different bands and ignore the strong constraints on
dynamics coming from the resonances.  The main purpose of this paper is to
check how well the models satisfy these resonance constraints when confronted
with the observational OGLE-2 Magellanic Cloud (MC) data.

\subsection{Models on the Tracks of Girardi \etal}

In Fig.~\ref{ber} we display sections of the evolutionary tracks of 
\cite{Girardi} for $Z=0.004$ and for $Z=0.008$ corresponding to the third
crossing of the instability strip (IS).  They are meant to be approximately
representative of the average properties of the SMC and LMC Cepheids.  We do
not show the second crossings of the IS here because they will give similar
results.

We note the well known fact that the blueward extents of the low mass
evolutionary tracks, \eg for 3.5\Mo\, are too short, and for the LMC the
3.0\Mo\ track does not even penetrate the IS.  This problem is not specific to
the Padova tracks, but is encountered by all recent evolution calculations.
See Cordier, Goupil \& Lebreton (2003) for a recent discussion of this problem
with evolutionary tracks and a possible explanation.  However the problem
occurs outside both the F and the O1 Cepheid resonance regions that we
are concerned with here.

As far as pulsation modelling is concerned the Padova evolutionary models
provide us with a mass $M$, a luminosity $L$, and effective temperature \Teff.
For a given composition (specified by $X$ and $Z$) these three quantities {\sl
uniquely specify the stellar envelope} \th \ie without a need to know the
properties of the stellar interior.  (One could equivalently, but much less
conveniently characterize the Cepheid envelope instead by the luminosity,
temperature and pressure at the core radius, for example).  We note that the
envelope composition is uniform having recently undergone a fully convective
stage.  The problems associated with core convection and overshooting (\eg
Cordier \etal\ 2003) thus need not concern us here because pulsation is limited
to the part of the star (the envelope) that is located above the burning shells
(\ie $T<$ a few million K).


 \begin{figure*}
 \centerline{\resizebox{16cm}{!}{\includegraphics{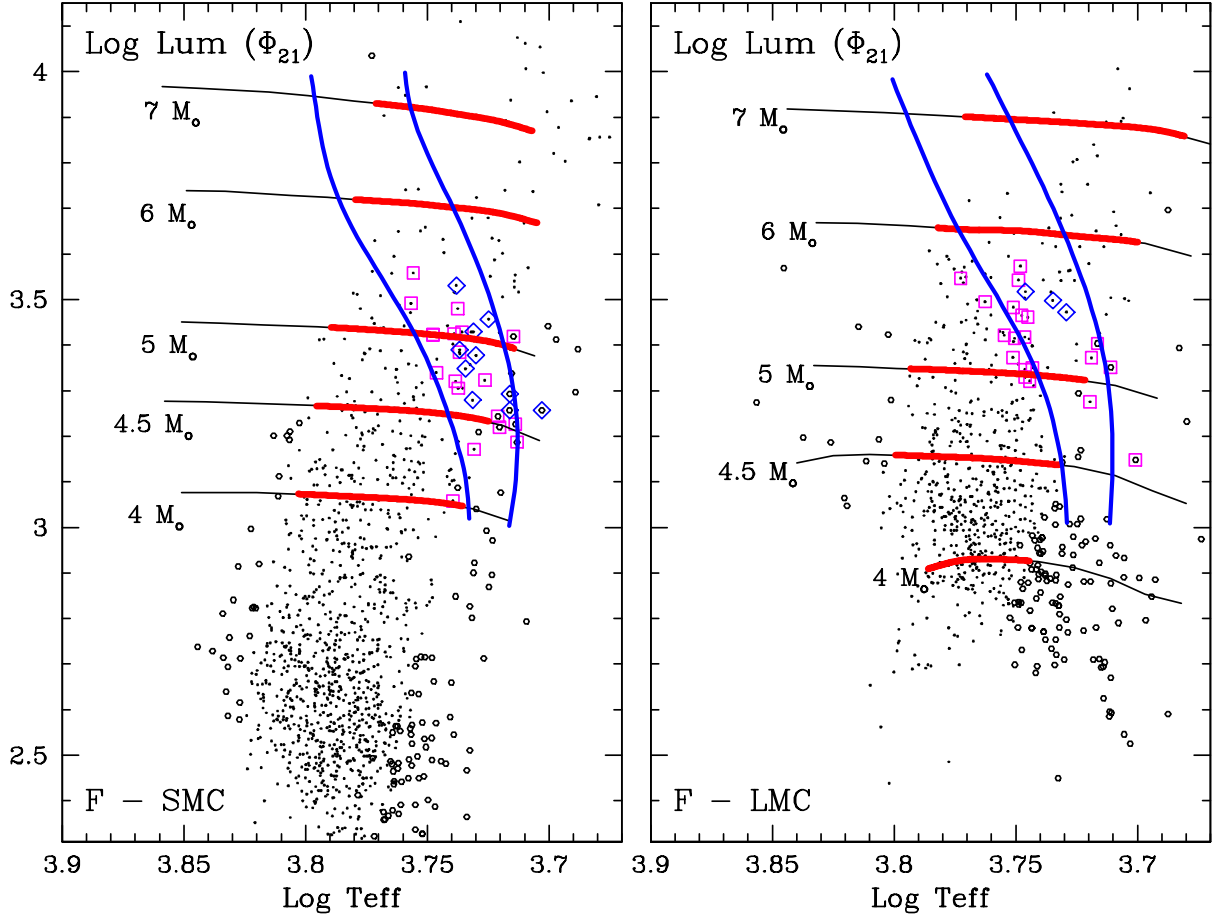}}}
 \vskip -1pt
 \parbox[b]{17cm}{
  \caption[]{\small {\bf MC F Cepheids -- $\phi^m_{21}$ resonance criterion:}
   {\sl dots (open circles)} represent linearly stable (unstable) OGLE models;
   {\sl squares}: $\phi^m_{21} > 5.45$ (and $P_0<20$), 
          {\sl tilted squares}: $0 < \phi^m_{21} < 2$.
The resonant Cepheid models lie between the vertical curves and the envelopes of
 the thickened tracks (instability strip).
            }
 \label{phiFB}
                 }
 \end{figure*}


 \begin{figure*}
 \centerline{\resizebox{16cm}{!}{\includegraphics{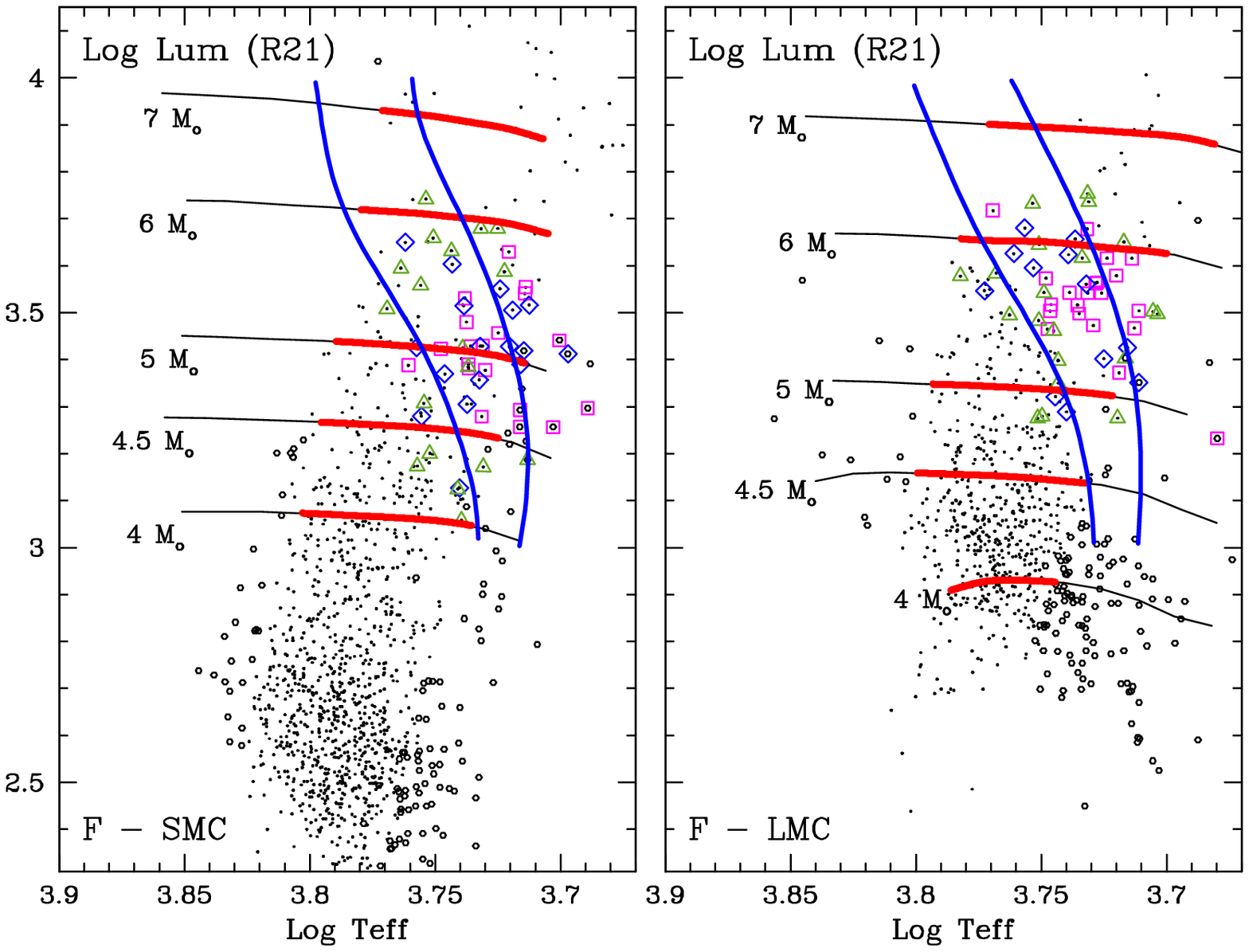}}}
 \vskip -1pt
 \parbox[b]{17cm}{
\caption[]{\small  {\bf MC F Cepheids -- $R^m_{21}$ resonance criterion:}
   {\sl dots (open circles)} represent linearly stable (unstable) OGLE models;
   {\sl squares}: $R^m_{21} < 0.15$ \th (and $P_0 < 16$), 
        {\sl tilted squares}: $0.15 < R^m_{21} < 0.2$ \th (and $P_0<16$),
    {\sl triangles}:  $0.2 < R^m_{21} < 0.25$ \th (and $6<P_0<16$).
    The resonant Cepheid models lie between the vertical curves and the envelopes of
 the thickened tracks (instability strip).}

 \label{r21FB}
 }
 \end{figure*}


The equilibrium Cepheid models and their {\sl linear} vibrational stability
properties are computed with our turbulent convective, TC code (\eg Yecko,
Koll\'ath \& Buchler 1998, Koll\'ath, Buchler, Szab\'o \& Csubry 2002) for the
model sequence along the Padova tracks.

One might think that there is a remaining inconsistency because Girardi
\etal~(2000) use standard (time-independent) mixing length in the envelope,
whereas we have additional ($\alpha$) parameters associated with our
time-dependent mixing length model, which also contains a turbulent flux and a
turbulent pressure.  The structure of the envelope turns out to be relatively
insensitive to these differences, as are the pulsation periods of the lowest
modes.  In contrast, as one may expect, the modal stability is more sensitive
to the $\alpha$'s.  We can therefore safely adopt the $M$, $L$ and \Teff\ given
by Girardi \etal\ and expect our envelopes to be very close to their
evolutionary ones.

The top plots in Fig.~\ref{ber} show the F Cepheid models and the bottom plots
the O1 Cepheid models.  The numbers above the tracks represent the linear
pulsation periods, $P_0$ and $P_1$ [d], respectively. \th (We recall here that
the nonlinear and linear pulsation periods are known to differ only by a few
tenths of a percent.)

The numbers below the tracks refer to the (linear) period ratios
$P_{r}=P_2/P_0$ and $P_{r}=P_4/P_1$, respectively, in days.  The pairs of thick
vertical curves delineate the boundaries of the resonance regions which we
define as $0.485<P_{r}<0.515$.

The thickened portions of the tracks in Fig.~\ref{ber} represent the linear IS,
\ie the region where the F or O1 Cepheid model, resp., are linearly unstable.
The actual IS is a little narrower because nonlinear effects are known to shift
the F blue edge a little lower in temperature, 0 -- 200\th K, depending on $M$,
and the O1 red edge a few hundred K to higher temperatures (\eg Fig.~2 of
Buchler 2000).  In addition, it is well known that there is a region where
either F or O1 pulsation is possible depending on the direction of the
evolutionary track (\eg Koll\'ath \etal\ 2002).


 \begin{figure*}
 \centerline{\resizebox{16cm}{!}{\includegraphics{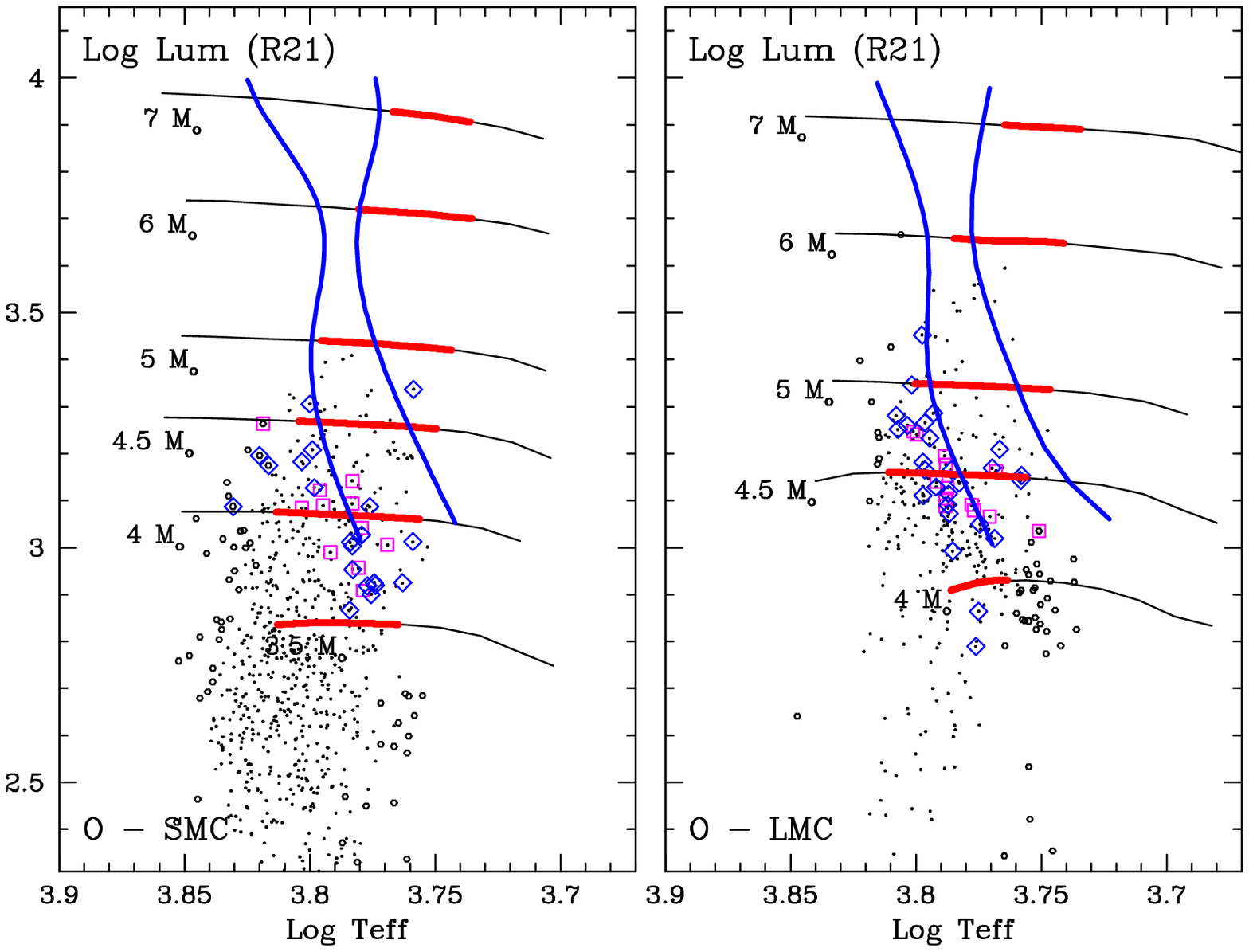}}}
 \vskip -1pt
 \parbox[b]{17cm}{
\caption[]{\small  {\bf MC O1 Cepheids -- $R^m_{21}$ resonance criterion:}
   {\sl squares}: $R^m_{21} < 0.05$ \th (and $P_1<4.5$), 
   {\sl tilted squares}: $0.05 < R^m_{21} < 0.07$ \th   (and $P_1<4.5$).
The resonant Cepheid models lie between the vertical curves and the envelopes of
 the thickened tracks (instability strip).
   The apparent lack of agreement is discussed in the text.
}
 \label{r21OB}
 }
 \end{figure*}


 \begin{figure*}
 \centerline{\resizebox{16cm}{!}{\includegraphics{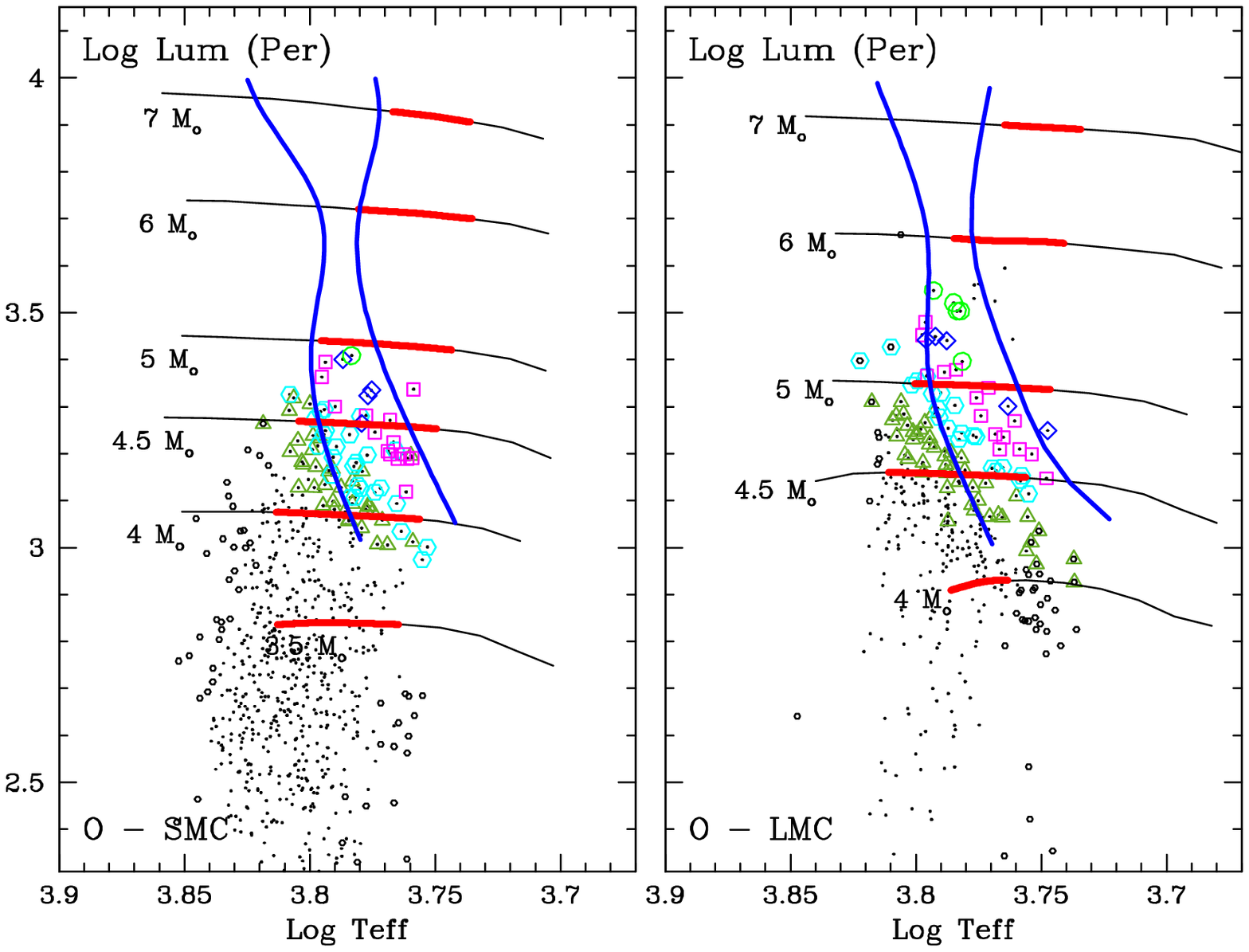}}}
 \vskip -1pt
 \parbox[b]{17cm}{
\caption[]{\small {\bf SMC O1 Cepheids -- period  'resonance' criterion:}  
{\sl dots (open circles)} represent linearly stable (unstable) OGLE models;
{\sl circles}:        $4.45 < P_1 < 4.95$, 
{\sl tilted squares}: $3.95 < P_1 < 4.45$, 
{\sl squares}:        $3.45 < P_1 < 3.95$, 
{\sl hexagons}:       $2.95 < P_1 < 3.45$,
{\sl triangles}:      $2.45 < P_1 < 2.95$.
The resonant Cepheid models lie between the vertical curves and the envelopes of
 the thickened tracks (instability strip).
 }
 \label{perOB}
 }
 \end{figure*}


In Fig.~\ref{ber} the resonant F Cepheids therefore should fall in the {\sl
resonance regions} defined by the resonance boundaries, $P_2/P_0 = 0.485$ and
$P_2/P_0 = 0.515$ (vertical curves) and the boundaries of the IS. \th (The
pulsationally unstable models marked by the thickened sections of the tracks).
These are the approximate values for which the (cos) Fourier coefficients
$R^m_{21}$ and $\phi^m_{21}$ have an egregious behavior, \viz the $R^m_{21}$
have a dip, and the phases $\phi^m_{21} \approxgt 5.5$ or $\phi^m_{21}
\approxlt 2.0$.

We see that for the resonant Cepheids, independently of $Z$ and of the
pulsation mode, the "NW" corner on the {\sl blue edge} of the IS, with the
highest $L$ and $M$ has the highest $P_r$.  For example for the SMC F Cepheids
one finds $M\sim 7.3$\Mo, with a period ($P_0\sim 16.5$d).  Conversely, the
"SE" corner of the {\sl red edge} of the resonant IS, with the lowest $L$ and
$M$ has the lowest $P_r$.  For the SMC F Cepheids $M\sim 4$\Mo with $P_0\sim
5$.  The resonant F Cepheid models along the $Z=0.004$ Padova tracks therefore
have periods ranging from $\sim 5$ to 16.5\th d.  A similar topography is seen
to obtain for the LMC Cepheids.

From the periods that are indicated in Fig.~\ref{ber} one can readily infer
that constant period curves run NW to SE, albeit with a slope that is shallower
than that of the constant $P_{r}$ curves.  If instead the resonance regions
were defined by the edges of the IS and the edges of the constant period
curves, they would be reasonably close to the resonance regions that we have
just defined on the basis of $P_r$, as one would expect from the correlation of
$R^m_{21}$ with the period in the OGLE data, for example.

A similar situation occurs for the overtone Cepheids where the resonant O1
Cepheid models along the $Z=0.004$ Padova tracks have periods ranging from
$\sim 2.0$ to 6.5\th d.

The constant $P_{r}$ curves that demarcate the resonant O1 models show a pinch
near 6\Mo.  Interestingly, this is not a numerical artifact, but can be traced
to the nonmonotone behavior of the the period $P_4$ of the fourth overtone,
because of the occurrence of a strange mode (Buchler \& Koll\'ath 2001, \eg
Fig.~2; Buchler, Yecko \& Koll\'ath, 1997).

For the $Z=0.008$ Padova tracks (third crossing of the IS) that are
approximately representative of the LMC Cepheids, we obtain a very similar
overall picture.  In particular, we find approximately the same period ranges
for the resonant F Cepheid models ($P_0 \sim$ 5.5 to 16\th d) and for the
resonant O1 Cepheid models ($P_1 \sim$ 2.2 to 7.2\th d).  Note that a similar
pinch in the resonance curves also occurs for the $Z=0.008$ tracks.

\subsection{Comparison with the OGLE-2 Cepheids}

\subsubsection{Instability Strip: Model Stability}

Beaulieu, Buchler \& Koll\'ath (2001, hereafter BBK) analyzed the SMC and LMC
OGLE Cepheids (OGLE-2 web database) with the intent of extracting $M$ -- $L$
relations from the observational data.  They first converted the observed
magnitudes and colors into $L$ and \Teff\ with assumptions about reddening and
distance moduli.  We shall refer to these observationally derived $L$, \Teff~
and periods as defining the {\sl OGLE stars}.  From these OGLE star data BBK
then derived stellar masses $M$ as well as linear stability properties with the
help of a linear pulsation code.  In the following we shall refer to these as
the {\sl OGLE models} when there is a need to distinguish between them.

We recall that BBK had concluded that observational luminosity and reddening
uncertainties cause a small spread in \Teff\ that is difficult to disentangle
from the spread in \Teff\ due to the width of the IS.  The derived $M$, $L$ and
\Teff\ values for the OGLE models thus contain small errors.  The two
consequences are that (1) in an $M$ -- $L$ diagram one obtains a swarm rather
than a narrow strip as one would expect from a relatively homogeneous group of
Cepheids, and that (2) in an \Teff\ -- $L$ plot some stars necessarily fall
outside the actual IS.

Figs.~\ref{phiFB} and \ref{r21FB} show theoretical HR diagrams (Log $L$ \vs Log
\Teff) in which we redisplay the OGLE F Cepheid stars for the BBK's (preferred)
choice B of distance modulus, namely 18.55$\pm$0.10 for the LMC and
19.97$\pm$0.15 for the SMC, and mean reddenings of E(B-V) = 0.1 and 0.08,
respectively.  These figures are identical except for the resonance criteria to
be addressed below.  In these figures we now represent the linearly
stable/unstable OGLE models with open circles/dots.  Despite the above
mentioned observational uncertainties in the derived $M$, $L$ and \Teff\ we see
that the majority of the OGLE Cepheid models are unstable, as they should be.
There appears to be some small systematic discrepancy, however, for the low
luminosity F Cepheids on the red side, both in the LMC and SMC that goes beyond
the uncertainties in the model parameters.  Considering that the linear growth
rates are the least certain of the calculated pulsation quantities, because
they depend on the $\alpha$ parameters that are used in the time-dependent
mixing length equations we are not too concerned because a small adjustment of
the $\alpha$ parameters in the convective terms could fix this problem.  We
stress again that, in contrast, the periods are largely independent of the
$\alpha$\th s.

In Figs.~\ref{r21OB} and \ref{perOB} we reproduce the OGLE O1 Cepheids.  One
notes that the vast majority of the OGLE models are linearly unstable, and that
the region occupied by the unstable evolutionary models coincides well with
that of the unstable OGLE models.

Overall, the figures thus indicate good agreement between the tracks, the
pulsation calculations and the observations as far as the stability of the
models is concerned.

\subsubsection{Resonance Conditions}

We now go on to examine how well the resonance information along the
evolutionary tracks agrees with that of the Fourier coefficients of the OGLE
LCs.  We start with the F Cepheids, both in SMC and LMC.

\vskip 5pt

 \underbar{\sl Fundamental Mode Cepheids:}

\vskip 5pt

As we have already discussed in the Introduction, for the observed F Cepheids
we know the periods, but we do not have any direct information on the {\sl
period ratios} $P_{r}$.  However, near $P_r\sim 0.50$ the period ratio
correlates with the dip of the LC's Fourier amplitude ratio $R^m_{21}$ .  It
also correlates well with the egregious values of the phases $\phi^m_{21}$ in
the resonance region, \viz $\approxlt 2.0$ and $\phi^m_{21} \approxgt 5.3$.
The phases are normalized to [0, 2$\pi$ ], mod 2$\pi$.

In Figs.~\ref{phiFB} and \ref{r21FB} we display \Teff\ -- $L$ plots for the F
Cepheids.  The OGLE F Cepheids are represented by dots for the linearly
unstable models, and by open circles for the stable ones.  Superposed are the
Padova tracks where again the thickened sections represent the linearly
unstable models (as in Fig.~\ref{ber}), and the vertical curves denote the
edges of the resonance regions, defined by $0.485 < P_r < 0.515$.

In Fig.~\ref{phiFB} we use the Fourier phase ($\phi^m_{21}$) resonance
criterion.  We have surrounded by {\sl squares} the stars for which
$\phi^m_{21} > 5.45$ (and $P_0<20$), and by {\sl tilted squares} those for
which $\phi^m_{21} < 2$.  The position of these resonant OGLE stars is seen to
agree rather well with the resonance region that we have defined with the help
of the models along the Padova tracks, despite the large scatter in the OGLE
Fourier coefficients that manifests itself in our figures.

In Fig.~\ref{r21FB} we use instead the resonance criterion based on the
$R^m_{21}$ coefficients.  We surround the stars for which $R^m_{21} < 0.15$ and
$P_0<16$ by square boxes, those for which $0.15 < R^m_{21} < 0.20 $ and
$P_0<16$ by tilted squares, and those for which $0.20 < R^m_{21} < 0.25 $ and
$6<P_0<16$ by triangles.  We have chosen these numerical values on the basis of
the plots of $R^m_{21}$ \vs period of the OGLE Cepheids (Udalski \etal 1999a,
1999b).

One notes that this second criterion gives results that are very similar to
those derived with the $\phi^m_{21}$ criterion.  Again there is excellent
agreement between the F Cepheid models along the Padova tracks and the OGLE F
Cepheids.

\vskip 5pt

\underbar{\sl First Overtone Cepheids:}

\vskip 5pt

We have already noted in the Introduction that for the O1 Cepheids neither the
phase $\phi^m_{21}$ nor the amplitude ratio $R^m_{21}$ are useful discriminants
for the location of the resonance (FBK).  Fig.~\ref{r21OB} shows that if we
persist in using $R^m_{21}$ as a resonance criterion, where the squares are
stars with $R^m_{21} < 0.05$ and $P_1<4.5$, and the tilted squares those with
$0.05 < R^m_{21} < 0.07 $ and $P_1<4.5$, we see that these 'resonant' stars
fall substantially below the resonance region defined by the evolutionary
tracks and pulsation theory.  The reason, as we have already pointed out, is
that the behavior of $R^m_{21}$ through the resonance is different (as a
function of $P_r$) for the O1 Cepheids and for the F Cepheids.  This point was
argued in FBK where it was shown that nonlinear Galactic overtone Cepheid
models give excellent agreement with the observational Fourier decomposition
data, but that the resonance occurs around 4.2\th d, rather than 3.2\th d.
Kienzle \etal~ (1999) had also suggested a resonance at 4.6\th d on the basis
of the Fourier decomposition of the $V_{r}$ curves of the Galactic overtone
Cepheids.

Since neither the $\phi^m_{21}$ nor the $R^m_{21}$ resonance criteria are very
simple or useful in the case of O1 Cepheids, we turn to the period $P_1$ as an
alternate resonance criterion in Fig.~\ref{perOB}, even though it is less
restrictive as pointed out in the Introduction.  The different period ranges
are indicated with different symbols, \viz circles: $4.45 < P_1 < 4.95$, tilted
squares: $3.95 < P_1 < 4.45$, squares: $3.45 < P_1 < 3.95$, hexagons: $2.95 <
P_1 < 3.45$, triangles: $2.45 < P_1 < 2.95$.

In the Introduction we suggested on the basis of the comparative location of
the minima of $R^m_{21}$ that in going from the Galaxy to the LMC and SMC one
might expect similar shifts of --0.5 and --0.7\th d in the locations of the
resonance regions.  The latter would thus be centered on $P_1\sim$ 3.7\th d for
the LMC (in the region of the squares in Fig.~\ref{perOB}) and on $\sim$ 3.5\th
d for the SMC (in the region straddling the squares and the hexagons).
Fig.~\ref{perOB} is certainly compatible with these conclusions.

\vskip 5pt

In this paper we have shown the properties of the OGLE Cepheids obtained with
BBK's choice B of distance modulus and reddening.  For completeness we have
repeated the same calculations with their choice A.  It is noteworthy that
choice B, which BBK labelled as preferred on other grounds, also gives better
agreement between the resonance properties of the observations and the models
than choice A.

\section{ Conclusions}

We have used the properties of the internal 2:1 resonance between fundamental
and second overtone modes for the F Cepheids (originally known as the
Hertzsprung progression), as well as the 2:1 resonance between the first and
fourth overtone modes for the O1 Cepheids to compare the observational, stellar
evolutionary and pulsational properties.  We find a very good agreement between
the evolutionary tracks of Girardi \etal\ (2000), our turbulent convective
pulsation code and the OGLE observational constraints.

The only major, and well known disagreement is the shortness of the blueward
evolutionary tracks for low masses where they do not even penetrate into the
instability strip, but this occurs outside the resonance regions.

Another, but smaller disagreement occurs in the precise location of the ISs and
their widths.  It has its origin partially in the errors in the $L$ and \Teff\
which themselves come from reddening and magnitude uncertainties in the
observational OGLE data.  From the modelling side uncertainties arise from the
values of the $\alpha$ parameters that enter the pulsation code through
time-dependent mixing length theory.  In addition, nonlinear effects that we
have not considered here shift the F blue edge to lower \Teff\ and the O1 blue
edge to higher \Teff.

In the past the sharp features in the Fourier coefficients, that are indicative
of the presence of internal resonances, have been used to localize internal
resonances.  This has worked reasonably well for the Galactic F Cepheids, but
as we have discussed here, in general, there exists {\sl no simple} \apriori
criterion for localizing precisely the resonance region from either the LC or
from the $V_{r}$ data.  One needs to resort to full amplitude numerical
hydrodynamic modelling to determine the criterion to be used to localize the
resonance center.  We find that we get much better agreement between the
resonant models and the OGLE data if we put the O1 Cepheid resonance center
($P_4/P_1$=1/2) at $\sim$ 3.7\th d for the LMC and at $\sim$ 3.5\th d for the
SMC, rather than where the $\phi_{21}$ has its sharp drop.  The same suggestion
was reached already by FBK in their detailed study of Galactic Cepheids where
this resonance appear $\sim$4.2\th d.


\section{Acknowledgements}
This work has been supported by NSF (AST03--07281) and Hungarian OTKA
(T-038440 and T-038437).

\vskip 10pt



\begin{thebibliography}{}

\bibitem[Antonello \& Aikawa (2000)]{AntonelloA}
    Aikawa, T. \& Antonello, E.,  2000, AA, 363, 601

\bibitem{ar}
 Alibert, Y., Baraffe, I., Hauschildt, P., \& Allard, F. 1999, AA, 344, 551

\bibitem[Antonello \& Poretti (1986)]{AntonelloP}
   Antonello, E. \& Poretti, E. 1986, AA, 169, 149

\bibitem[Antonello \& Poretti (1986)]{AntonelloP}
   Antonello, E.,  1994, AA, 282, 835

\bibitem[Antonello \& Aikawa (1995)]{AntonelloA}
    Antonello, E. \& Aikawa, T., 1995, AA, 302, 105

\bibitem[Antonello \& Aikawa (2000)]{AntonelloA}
    Antonello, E. \& Aikawa, T., 2000, AA, 363, 601

\bibitem{ar}
Baraffe, I., Alibert, Y., M'era, D., Chabrier, G., \& Beaulieu, J. 1998, ApJ, 499, L205

\bibitem{BBK}
Beaulieu, J.P. Buchler, J. R. \& Koll\'ath, Z., 2001, AA 373, 164 [BBK]

\bibitem{a}
 Beaulieu, J.P., Grison, P., Tobin, W.  et al., 1995, AA 303, 137

\bibitem{a}
 Beaulieu, J.P. \& Sasselov, D. 1997,
 in ``Astrophysical Fallouts From Microlensing Projects'',
 Eds. R. Ferlet \,J.P. Maillart \& B. Raban, Editions Fronti\`eres,
 Gif-s-Yvette\th  (also astro-ph/9612217)


\bibitem[]{Bono}
Bono, G., Marconi, M., \& Stellingwerf, R.F., 2000a, AA 360, 245

\bibitem[]{Bono}
Bono, G., Caputo, F., Cassisi, S., et al. 2000b, ApJ, 543, 955

\bibitem[]{Bono} Bono, G., Gieren, W. P., Marconi, M., Fouqu\'e, P., \& Caputo,
F. 2001, ApJ, 563, 319

\bibitem[Buchler (1993)]{BuchlerMito}
  Buchler, J. R. 1993, in {\it Nonlinear Phenomena in
  Stellar Variability},Eds. M. Takeuti \& J.R. Buchler (Kluwer: Dordrecht),
  repr.~from ApSS 210, 1

\bibitem[Buchler \& Goupil (1984)]{BuchlerG}
  Buchler, J. R. \& Goupil, M. J. 1984, ApJ 279, 394

\bibitem{s} Buchler, J. R. \& Koll\'ath, Z., 2001,  ApJ 555, 961

\bibitem{}
Buchler, J.R., Koll\'ath, Z., Beaulieu, J.P. \& Goupil. M.J.  1996,
    ApJLett, L462, 83

\bibitem{}
  Buchler, J. R. \& Kov\'acs, G. 1986, ApJ, 303, 749 

\bibitem[Buchler, Moskalik, \& Kov\'acs (1990)]{BuchlerMK}
  Buchler, J. R., Moskalik, P. \& Kov\'acs, G. 1990, ApJ 351, 617

\bibitem{d} Buchler, J.R., Yecko, P.E. \& Koll\'ath, Z. 1997, AA 326, 669

\bibitem{cordier}
  Cordier, D., Goupil, M. J. \& Lebreton, Y. 2003, AA 409,491

\bibitem[Feuchtinger, Buchler \& Koll\'ath 2000]{FeuchtingerBK} 
 Feuchtinger, M., Buchler, J. R. \& Koll\'ath, Z., 2000, 
 ApJ 544, 1056 [FBK]

\bibitem[Girardi \etal, (2000)]{Girardi} 
 Girardi, L., Bressan, A., Bertelli, G. \& Chiosi, C. 2000, AAS 141, 371

\bibitem[Hertzsprung]{Hz}
 Hertzsprung E. 1926, Bull. Astr. Inst. Netherlands 3, 115

\bibitem[Kienzle \etal (1999)]{Kienzle}
  Kienzle, F., Moskalik, P., Bersier, D., Pont, F. 1999, AA 341, 818

\bibitem{Klapp}
 Klapp, J., Goupil, M.-J. \& Buchler, J. R. 1985, ApJ, 296, 514

\bibitem{KBSC}
  Koll\'ath, Z., Buchler, J.R., Szab\'o, R., Csubry, Z. 2002 ApJ 573, 324

\bibitem[Kov\'acs \& Buchler (1989)]{KovacsB89}
  Kov\'acs, G., Buchler, J.R., 1989, ApJ, 346, 898

\bibitem[Kov\'acs \& Buchler (1990)]{KovacsB90}
  Kov\'acs, G., Buchler, J.R. \& Moskalik, P. 1990, ApJ, 351, 617

\bibitem[kkb]{kkb} 
  Kov\'acs G., Kisvarsanyi E. \& Buchler J.R. 1990, ApJ  351, 606

\bibitem[MBM]{}
Moskalik, P., Buchler, J.R. \& Marom, M. 1992,  ApJ  385, 685

\bibitem[Payne]{PG} 
  Payne-Gaposhkin, C. 1947, AJ, 52, 218

\bibitem[Simon \& Kanbur (1994)]{SimonK}
  Simon, N.R. \&  Kanbur S. 1994, ApJ, 429, 772

\bibitem[Simon \& Schmidt (1976)]{SimonS}
  Simon, N. R. \& Schmidt, E. G. 1976, ApJ, 205, 162

\bibitem[]{}
   Udalski, A., Soszynski, I., Szymanski, M. et al., 1999a, AcA 49, 223

\bibitem[]{}
   Udalski, A., Soszynski, I., Szymanski, M. et al., 1999b, AcA 49, 437

\bibitem[Yecko, P.A., Koll\'ath, Z., Buchler (1998)]{YeckoKB} 
  Yecko, P.A., Koll\'ath, Z., Buchler, J.R. 1998, AA 336, 553


\end{thebibliography}
\end{document}